\documentstyle[Headings,11pt]{article}

\bottomarque{Success and failure of programming environments}

\date{Jan 1990}
\begin{document}
\thanks{This work was supported by the CEE (Esprit Project STAPLE) and
  by Non Standard Logics, Paris, France}
\author{C. Recanati}
\title{Success and failure of programming environments\\
-  Report on the design and use of a graphic\\
 abstract syntax tree editor -}
\maketitle
\begin{abstract}

  The STAPLE project investigates a persistent architecture for functional
  programming. Work has been done in two directions: the development of a
  programming environment for a functional language within a persistent system
  and an experiment on transferring the expertise of functional prototyping
  into industry. This paper is essentially a report on the first activity.\\

  The first section gives a general description of Absynte - the abstract
  syntax tree editor developed within the Project. Following sections make an
  attempt at measuring the effectiveness of such an editor and discuss the
  problems raised by structured syntax editing - specially environments based
  on abstract syntax trees. Although the benefits of syntax directed editors
  are obvious for beginners, the conclusion is that they are not very
  attractive for experimented users. \\

\end{abstract}

\newpage

\section{Absynte: a graphic abstract syntax tree editor}

\subsection{Description of the editor}

  The Staple editor is called {\it Absynte} where these letters stand for {\bf
  Ab}stract {\bf syn}tax {\bf t}ree {\bf e}ditor. Absynte allows the edition of
  programs stored in a persistent database as decorated abstract syntax trees
  (ASTs). A program can be displayed as several views: graphics views, text
  views and views mixing text and graphics together. Absynte also gives access
  to other tools manipulating the programs stored in the database, such as a
  {\tt compile\_and\_run} command.\\

  Let us begin by an informal description of how a program creation is handled
  within the Absynte environment. First of all, absynte is a program running
  under the X-window System - which has been welcomed as a universal low-level,
  portable graphic and windowing system. Absynte starts by opening a window
  decorated with a title bar that distinguishes Absynte graphic windows from
  other windows on the screen. \\

  At the top of the window a menu bar provides a number of items in pop up
  menus: {\tt Store}, {\tt File}, {\tt Edit}, {\tt Window}, {\tt Layout} and
  {\tt Mode}. A bottom bar gives the name of the program being edited and
  some additional information. As an example, opening the File menu and
  launching the {\tt new} command will put the editor in creation mode. Now
  if the user drags the mouse pointer to the center of the window and click, a
  first node containing the string {\tt prog} appears.\\

  Since the node labeled {\tt prog} is incomplete, the user can click on it.
  This will extend the {\tt prog} node with two sons: an optional list of
  definitions labeled {\tt define*}, and an optional {\tt expression*} to
  evaluate.
  Since it is itself incomplete, the {\tt expression} node can be expanded to
  another tree. A click on it will show the list of all possible ways of
  completing it (see figure \ref{completion} below).
\begin{figure}[htbp]
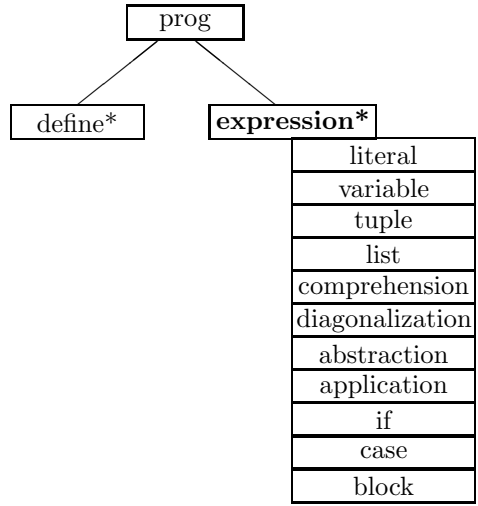

\begin{center}
\include{completfig}
\end{center}
\caption{A completion menu}
  \label{completion}
\end{figure}

  The user can then select an item by clicking on it. This will develop the
  selected subtree by replacing the incomplete typed node by a precise
  instance. Otherwise, clicking outside the menu will cancel the extension
  procedure. \\

  At some point a terminal definition will be reached. For instance, if the
  user has chosen a {\tt literal} in the {\tt expression} menu, and then
  selected
  an {\tt integer} value in the next submenu, he will be offered no further
  choices; instead, a little window will prompt him to enter a terminal
  string.\\

  At any time new windows can be created and their content edited in any mode.
  The user can load programs from the persistent store or create new ones. He
  can
  cut or copy trees or subtrees selected in any window in a cut buffer. The
  cut buffer is not attached to a particular window but is shared by all
  graphic windows. The {\bf Paste} and {\bf Replace} commands are used to
  paste the contents of the shared buffer into the window from whose menu bar
  they have been activated. If there is a type mismatch, no replacement will
  occur and the editor will protest.\\

  There are several ways of modifying the view of a tree. First you can edit a
  specific node in graphic mode or in text mode \footnote{In the standard
  Staple environment, the text mode uses a Miranda-like concrete syntax, but
  the concrete syntax can be changed at will.}.
  The natural situation is to
  have the whole program in graphic mode and some nodes displayed as text in
  separated windows. It is also possible to have unexpanded nodes, whose icons
  are associated with separated graphic windows.\\

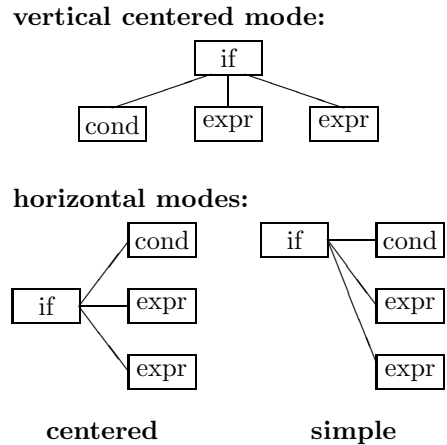
\begin{figure}[htbp]
\begin{center}
\setlength{\unitlength}{0.11mm}
\begin{picture}(560,560)(-680,700)
\put(-520,940){\framebox(80,40){\small  cond}}
\put(-660,1000){\small\bf horizontal  modes:}
\put(-580,880){\line(1,0){60}}
\put(-580,880){\line(3,4){60}}
\put(-520,860){\framebox(80,40){\small  expr}}
\put(-660,860){\framebox(80,40){\small  if}}
\put(-360,940){\framebox(80,40){\small  if}}
\put(-280,960){\line(3,-4){60}}
\put(-280,960){\line(1,0){60}}
\put(-600,780){}
\put(-520,780){\framebox(80,40){\small  expr}}
\put(-580,880){\line(3,-4){60}}
\put(-220,780){\framebox(80,40){\small  expr}}
\put(-280,960){\line(2,-5){60}}
\put(-220,860){\framebox(80,40){\small  expr}}
\put(-220,940){\framebox(80,40){\small  cond}}
\put(-620,720){\small\bf centered}
\put(-300,720){\small\bf simple}
\put(-440,1080){\framebox(80,40){\small  expr}}
\put(360,-120){\begin{picture}(0,0)
\end{picture}}
\put(360,640){}
\put(260,-560){}
\put(540,500){}
\put(500,660){}
\put(300,-360){}
\put(40,-240){\framebox(0,0){}}
\put(-460,1380){}
\put(-20,1060){\begin{picture}(40,0)
\end{picture}}
\put(-660,1220){\small\bf vertical centered mode:}
\put(-640,460){}
\put(-280,1040){}
\put(-300,1080){\framebox(80,40){\small  expr}}
\put(-580,1080){\framebox(80,40){\small  cond}}
\put(-380,1160){\line(3,-1){120}}
\put(-540,1120){\line(3,1){120}}
\put(-400,1120){\line(0,1){40}}
\put(-440,1160){\framebox(80,40){\small  if}}
\put(-380,1140){}
\put(-280,860){\begin{picture}(0,60)
\end{picture}}
\end{picture}
\end{center}
\caption{Graphic modes}
  \label{screen}
\end{figure}

  Within the graphic mode the user can choose between different styles
  in a window: vertical centered mode, horizontal centered mode and simple
  horizontal mode (see figure \ref{screen}).  The distance
  between graphic nodes can be changed interactively. A node
  and all its
  descendants can also be hiden and replaced by a graphic icon; a
  window will
  automatically be associated with this icon. Clicking on the icon displays
  the corresponding window which shows the corresponding subtree as it was
  originally in the parent window (this is recursive procedure and you can
  have iconified nodes in it). At any time iconified nodes can be expanded in
  two ways: global expand (recursive) or simple expand.\\

  To display a node in text mode, select the command {\tt Text Mode}. This
  will iconify the selected node (and all his descendants) as a text icon and
  display the tree as text in an associated window. An icon corresponding to a
  text window will appear with a capitalized T on it. Clicking on a text icon
  with the left button opens the corresponding text window. A new click on the text icon
  will close it. Clicking on a
  text icon with the middle button  opens a menu containing the following
  commands: {\tt Parse text}, {\tt Graphic tree}, {\tt Show window} and {\tt
  Hide window}.\\

  The {\tt Parse} command launches a parser checking the syntax of the text
  being edited. Error messages will be displayed if the text does not fit the
  concrete syntax. The {\tt Graphic} tree command moreover performs graphic
  conversion of the text contained in the text window. It will put the tree
  back in the main window as a graphic tree and remove the text icon and the
  associated text window. \\

  A window in text mode provides a full text editor augmented with syntax
  facilities. There are many commands in text mode because Absynte has used a
  very powerful text server \footnote{WX - a X-window general text editor
  developed independently at N.S.L.}.
  In particular, the user can easily move the cursor around in the text by
  means
  of the mouse pointer, he can search and replace segments of text in various
  ways, select one font per text window and load fonts on line as well.
  Furthermore certain keywords are recognized by the editor, and, when
  followed by an extension key (for instance the tab character), a template is
  automatically inserted. This syntax feature has been implemented through the
  definition of aliases and is not directly related to the ASTs structures. \\

  The session can be finished by launching the {\tt Save} command of the Store
  menu. This opens a dialogue box. A default name is proposed in a text area
  but it is editable. Clicking on the Confirm button or pressing the return
  key, will save the tree in the persistent store under the given name. The
  save can be canceled by clicking on the Cancel button. There are also two
  other saving procedures available in the editor: the user can save (and
  load) text from the Unix file system in the text editor mode. Graphic trees
  can also be saved (and loaded) as unix files by using the File menu of the
  main window\footnote{This serves as an interface to the file systems and in
  addition allows the use of Absynte outside a persistent environment.}.\\

  Saved programs can be compiled and run from Absynte. The execution appears
  in a separated window. At the end of execution, this window is deleted
  by pressing the return key.\\

\subsection{The obvious benefits of an internal tree structure}

  In the Staple system abstract syntax trees are the internal structures of
  programs. In this section we enumerate the obvious benefits of the use of
  AST in our system.

\subsubsection{A clear syntax}

  As far as user interface is concerned, AST graphs give a very clear
  representation of the program structure. It removes all ambiguities possibly
  introduced by some specific features of the concrete syntax. It is
  fundamentally not ambiguous (no {\it dangling else} problem), and suppresses,
  for instance, the need for parenthesis. Furthermore the Absynte graphic view
  perfectly reflects the internal representation of the program and is a good
  frame for the top-down creation of a program - which, as we have seen, can be
  performed by clicking on incomplete nodes and selecting nodes in completion
  menus. \\

  Many syntax editors have chosen ASTs as their internal representation of
  programs. This choice is independent of the external representation, in our
  case a graphic tree in graphic windows, and for instance in the Centaur
  system, a text representation. What is usually noted  in the literature of
  systems based on ASTs is their
  use of templates, which are very easy to provide \footnote{
In our case, we do not properly provide templates but
  graphic nodes stand themselves for templates and our system shares the
  properties
  (positive and negative) usually associated with template models.}. In the 
  next section, we will discuss the
  properties of the so-called template model. Its main benefits are:
  \begin{itemize}
\item typing effort is reduced, and the possibility of introducing
  typing error minimized.
\item generated programs are syntactically correct.
\end{itemize}

\subsubsection{A frame for the implementation of Persistence}

  Another strong feature of ASTs is that they provide a clear framework for
  the implementation of persistence. Persistence is a major concept of the
  Staple project. The persistence of data is the length of time for which it
  exist. One usually distinguishes between long-lived data structures
  (traditionally restricted to a file or database management system
  organization) and short-term data structures (normally expressed with
  programming language structuring facilities).\\

  In traditional programming there often is a considerable amount of code,
  typically 30 percent of the total, concerned with transferring data to and
  from files or a database management system. Much space and time is taken up
  by code to perform translations between the program's form of data and the
  form used for the long term storage medium. Thus in the program development
  process a lot of time is used to transfer data from one tool to another.\\

  The advantages of orthogonal persistence in a programming system are numerous
  and widely documented (see \cite{Atkinson85} and \cite{Atkinson86}). A
  design is called {\it orthogonal} if all components can be mixed freely,
  without any restriction. Orthogonal persistence makes the persistence of a
  data object independent of how the program manipulates that data object.
  Another advantage is the removal of discontinuities in the design of software
  systems and their user interfaces. This is also a goal of a
  programming environment designer. Thus persistent systems provide an ideal
  base technology for integrated programming environment.\\

  Traditionally, integrated programming environments have used a succession of
  files to represent the various stages of the software in development - from
  text to run-time structure. This complexity is unnecessary within a
  persistent system. It collects the various program representations - text,
  abstract trees, abstract trees decorated with various semantic information
  (comments, codes, graphic annotations or whatever) - and puts them together
  in the persistent store as multiple representations of the same object. All
  tools of a persistent system such as compilers, editors, interpreters,
  debuggers etc. can access and use the same information. In this way, much
  time and space is saved and coherence problems are easier to solve.\\

  The language used in the Staple programming environment is a typed
  functional language, very similar to Miranda. Functional programming also
  contributes to improve software productivity in several areas. Firstly,
  functional programs tend to be shorter than those written in other
  languages. Secondly, functional languages are good for prototyping as
  compared with other specification techniques. They have the great advantage
  of not having to specify the control flow of the program. This removes an
  important disturbing element from the activity of programming; another good
  feature is the flexibility of the notations allowed.\\

  Some other advantages of the Staple language are
  laziness, orthogonality and polymorphism. Laziness simplifies programmming.
  It suppresses tests and the need of secondary
  functions (for instance the use of iterators in unification programs). It
  also allows the definition of streams - which are very useful for modeling.
  Orthogonality and polymorphism also simplify the writing of the program.
  Polymorphism allows the definition of generic operations which work on
  different types of data, in such a way that functions need not be rewritten
  for each data type in the abstract function's domain.\\

  Some functional languages already provide module systems and separate
  compilation ({\tt HOPE, Miranda, Haskell}). However these modules are flat
  code rather than objects. Thus the evaluation of a module needs to be
  repeated during the execution of every program which imports that object from
  the module. The next step consists in treating modules as permanent objects
  of a persistent store; this was an important motivation of the Staple
  project.\\

  The programmer now has the opportunity to use data types and functions from
  the persistent store rather than having to define every entity which she
  uses. The persistent store thus provides a programmer with a library of
  defined entities. In contrast to the inclusions in a program of previously
  typed text which has been saved in a file and must be recompiled before
  execution, the use of a function which had been saved in the persistent
  store is a run-time operation, with the evaluator obtaining the code for the
  function from the persistent store.\\

\subsubsection{A frame for the integration of tools}

  ASTs provide a general framework for the integration of the various tools
  using the program. For instance, the editor can highlight errors found by the
  compiler within the text and move on user request to the next error to be
  corrected. Here a persistent programming environment is helpful because the
  compiler can decorated the abstract tree with appropriate annotations which
  the editor may access.\\

  As well, the abstract syntax tree representation is useful for the building
  of a debugger. The reason is the same as in the compiler case: the abstract
  tree can be decorated with piece of code and then the execution of a program
  can be displayed as a traversal of the program tree.\\

\subsubsection{A frame for language specifications}

  At the beginning of the project all partners agreed on the fact that the
  language should be functional, lazy, strongly typed and provide fundamental
  pattern-matching features. We then isolated an abstract syntax. But it was
  very difficult to find an agreement on concrete syntax. Then Absynte was
  designed so as to generate a syntax directed editor from a list of
  specification rules. The generator system uses the specification of a
  language as input and produces as output the required syntax-directed
  editor. Abstract syntax trees were already the basis of our implementation
  of persistence, and they provide an easy frame for language specifications.
  The specification rules were designed to define the ASTs set and also 
  to give some instructions to the system for driving the editor. Then we were
  free to change the abstract syntax while developing the user interface. \\

  ASTs are made available to the generator system by means of a formalism
  allowing the definition of class rules and production rules. The class rules
  define the syntactical categories of nodes. The production rules describe
  the structure of a node by defining the number and the types of its sons.
  The class rules are used by the generator system to create menus for the
  expansion of incomplete nodes and the production rules are used for
  displaying the nodes. Lately, ASTs (as production rules)
  have also been decorated with pretty-printing annotations allowing
  automatic translations from graphic representation to text representation.\\

  Today the generator system produces an abstract syntax tree editor for an
  arbitrary abstract syntax but not a complete environment (with compiler and
  interpreter) because we do not have implemented {\it semantic}
  specifications. The Staple environment includes a compiler and an
  interpreter, but these tools are specific to the functional language used by
  Staple and not generated by the parsing of semantic specifications.

\section{Syntax directed editors: some problems}


  In the preceding section we have
  enumerated the advantages of the use of ASTs in our syntax directed editor.
  To summarize, they generally provide the following benefits:
\begin{itemize}
\item ASTs remove contingent ambiguities introduced by concrete syntax.
\item ASTs provide a good implementation structure in many respects: trees are
  easy to manipulate and can be decorated at will. 
\item ASTs facilitate the automatic generation of syntax-directed editors 
  by providing a frame for syntax specifications.
\item ASTs lead to the generation of
  syntax editor based on the {\it template model}.
\end{itemize}

  In this section, we will be more critical towards the practical use of ASTs in
  syntax
  directed editors. 

\subsection{Template and token model: top-down vs bottom-up}

%
  The major concept of the template model is a template containing {\it place
  holders}. These place holders can be expanded by means of other templates;
  this is often supported by commands or menus to choose a permitted
  construct. Text must be inserted if the place holder can no longer be
  expanded: the only allowed construct is then a terminal symbol.\\

  The text always remains syntactically correct, since the type of permitted
  construct is determined by the type of the place holder. A type checker is
  nevertheless often used to perform additional semantic checking.\\

  In the token model, text editing is permitted everywhere. The text is checked
  continuously during editing with an incremental parser. The user
  knows at each moment whether the text
  he has typed is correct or not.\\

  The template model and the token model can also be viewed as top-down model
  and bottom-up model. The template model is a top-down model since the
  program is created from a top template by inserting allowed constructs. A
  top-down strategy is well suited to the creation of a new program. With
  expansion of templates a new program can be quickly constructed and is
  guaranteed to be syntactically correct. This feature is especially useful for
  novices. \\

  A first problem within a pure template model is that only terminal symbols
  can be entered as free text. This characteristic is absolutely intolerable in
  practice. Suppose you want to enter an arithmetic expression containing
  three or four operators. Then you should select the appropriate template by
  means of commands or menus for each operator and argument - which results in
  many additional commands being triggered.\\

  This is why a pure template model has never been implemented in practice.
  Both the Cornell Program Synthesizer (\cite{Reps84}, \cite{Teitelbaum}) and
  Mentor (\cite{Donz80}) have used an hybrid model of top-down replacement of
  place holders with template and bottom-up construction of phrases from free
  text. We have also adopted such a model in Absynte. In a hybrid model, a
  focus is used to delimit an area of free text editing. In this area the text
  may be temporary incorrect. In the systems where the internal representations
  are ASTs, a first drawback is the administration required for parsing and
  unparsing the new representation.\\

  If a free area is really useful to enter arithmetic expressions it does not
  solve the most important defect of the template model: how to
  handle modifications within a tree structure ?
  The replacement of a {\it while}
  statement by an {\it until} or {\it if} statement is more difficult than the
  corresponding change in a traditional text editor (the later is quite
  simple). In general, the replacement of one construct by another causes all
  information belonging to the old construct to be deleted. If the user wants
  to save some pieces, he can sometimes do it in special buffers or windows,
  but he must perform these operations separately (for each syntactic category
  and, of course, before the complete removal) because every piece of program
  to be saved must be inserted in its future template separately. \\

  Hybrid models do not solve this problem. When the text can
  be freely introduced in a free area, it must nevertheless be at some point
  converted into an abstract syntax tree node. Thus, although free text is
  allowed in a free area, not all expression can be in practice introduced
  there and the user must take care of the syntactical types. This constraint
  demands some particular attention and complicates the editing task instead of
  simplifying it. \\

  The fact that no pure template model exists only shows the obvious
  superiority of the token model. Note that the difference between the token
  based model and the template model is not that the representation provided by
  the system is textual in one case but not in the other. Most of systems
  based on a template model have used text for displaying abstract trees. The
  true difference is the way a program text can be entered. Template models
  are plagued by the problem of program modifications, even if the
  representation provided is textual.\\

  Modifications such as "while-to-until" are awkward because they are neither
  considered nor specified in the language definition. The only way to solve
  this problem is to have an extensional list of all possible tree
  transformations allowed. Each valid editing operation should be designed as
  a tree transformation rule specifying when and how this editing function may
  be applied. In other words, the only way of handling tree modifications is
  to define a specification language specially designed for this purpose! This
  solution as been taken seriously by Arefi, Hughes and Workman for the
  automatic generation of visual syntax-directed editors (\cite{Arefi90}).
  Although it certainly does not satisfies the Occam razor principle, it is
  the only solution to the building of user-friendly graphic
  syntax-directed editors.\\

  The properties of the token model are exactly opposed to the ones of the
  template model. In the token model
  text editing is permitted everywhere in the window and it is a
  bottom-up editing model. Text is perceived as a sequence of tokens, which
  means that each word in the text has a lexical type assigned to it. Because
  of the text oriented character of the token model, there is generally no
  possibility for manipulating language constructs. \\

  If changes are easy, parsing is difficult and program creation is harder
  than with template models. Nevertheless some system provide automatic
  correction (as variable declaration insertion in Cope \cite{Archer81}). The
  text must be parsed to check errors after each modification. The drawback
  from the user's point of view is that many errors may be permanently
  highlighted because the writing of a program includes a lot of incorrect
  temporary states. From the viewpoint of the designer the drawback is that an
  incremental parser \footnote{Incremental parsing is difficult to implement.
  To be efficient, it must minimize the amount of parsing by
  comparing the old and modified text, which requires a lot of administration.}
  is called for.  \\

  Very few systems have integrated a token model based editor. Interesting work
  has been performed in automatic generation of bottom-up incremental parsers
  from language specifications in the Gipe project (\cite{Heering88a},
  \cite{Heering88b}) but the text editor of the current Centaur system is
  still based on an hybrid model and the parsing command must be selected
  in a menu to parse the content of a free text area. The same criticism can be
  made to Absynte.

\subsection{Multi-language system problems}

  We can also divide existing systems into different categories according to
  their functions: some are specially built for a specific language
  while others are multi-language systems. Pascal has frequently
  been chosen as a specific language (Omega, Poe, Magpie and Pases) because of
  its block structure and its simplicity. By contrast, the Cornell Program
  Synthesizer, Pecan, Centaur and Staple are examples of multi-language
  systems.\\

  Systems built for a specific language have great advantages. They
  are more simple and therefore easier to implement. But the most important
  thing is that they can take into account any
  particular characteristic of the language. \\

  Yet most of the systems appear to have evolved from the production of a
  single programming environment for a specific language to the generation of
  other environments from language specifications. For instance, both the
  Cornell program synthesizer and the Mentor system were in their first version
  designed for a specific language. \\

  We think that this general move has been an interesting experiment, but
  failed at producing pragmatic programming environments. The design of these
  systems was based on the idea that by means of abstract descriptions, the
  generation of syntax-directed editors would not be too expensive. But this
  is untrue\footnote{Firstly because this generalization is expensive.
  Secondly, it is not really effective. In all actual systems, the term
  'multi-language' is a bit too strong. In practice, the languages that can be
  introduced are all of the same family - imperative or functional - and such
  systems are not prepared to the automatic generation of a good C++ (object
  oriented) language editor. Furthermore even in traditional programming the
  generated environment is in general not efficient enough to satisfy the
  user.}.\\

  In this section, we are going to give a few examples taken from various
  systems to illustrate this point and give the reader the flavor of
  language specification problems.

\subsubsection{Syntax specifications}

  Syntax specifications are the starting point of the generation of specific
  programming environments. It includes abstract and concrete syntax
  specification and, for some system such as Centaur and Staple,
  pretty-printing rule specifications. Additional information is frequently
  required by the system to show other views of the program or to define the
  relationships between abstract specifications and concrete specifications.
  In systems using ASTs, the compiler for these specifications usually
  provides a parser and an unparser from AST to text and reverse.\\

  Two typical approaches are given by Pecan and Centaur.
  Both are interesting in the way the specifications of abstract and
  concrete syntax are given. It shows in particular that the relationships
  between these descriptions is the cornerstone of such
  systems.\\

  In Centaur, the  language in which concrete and abstract syntax are specified
  together with their relationships is called METAL (see \cite{Borras88b}).
  A Metal specification of a formalism F (a language to be added to the system)
  consists of three parts:
\begin{itemize}
	\item the definition of the concrete syntax in terms of BNF-like
  rules: this is a set of rules that make it possible to decide whether or not
  a given sentence belongs to the formalism. These rules will be used to
  construct a parser for F.
	\item the definition of the abstract syntax in terms of {\it
  operator} and {\it phyla}: to define the set of correct abstract syntax
  trees.
	\item a list of tree building functions: these functions specify the
  connection between the concrete syntax and the abstract syntax.
\end{itemize}
  The compilation of this specification produces a parser and a tree
  generator. \\

  In Pecan, the specification uses a statement-oriented semantic language
  (\cite{Reiss846}). Each abstract syntax construct is annotated with the
  information needed by the system to build and define various views. An
  example of specification for the Pascal {\it while} statement
  is given in figure \ref{whilefig}.\\
\begin{figure}[htbp]
\begin{verbatim}
while_statement=>EXPRESSION STATEMENT ::
 SOURCE: "WHILE @1 DO @+@R@c@2@_@R@C"
 COMMENT
 SYNONYM: "While"
 NS: LOOP @1 @2 NONE;
 SEMANTICS: {
   BEGIN loop;
    DEFINE NAME=operator,EXIT,CLASS=label;
    DEFINE NAME=operator,NEXT,CLASS=label;
    USE NAME=operator,NEXT,CURRENT=ONLY;
    FLOW LABEL=1,LABEL=REF;
    DO @1;
    FLOW NOTTEST,2;
    DO @2;
    FLOW GOTO=1;
    USE NAME=operator,EXIT,CURRENT=ONLY;
    FLOW LABEL=2,LABEL=REF;
   END;
   };
\end{verbatim}
\caption{Annotated Abstract Syntax for the WHILE Statement} \label{whilefig}
\end{figure}

  The main advantage of the Pecan approach is its simplicity . The same
  description specifies syntactic and semantic aspects as well as
  pretty-printing rules. Consequently a modification requires only one change
  to the structure concerned. We have adopted a similar method in Staple. The
  relationships between abstract syntax, concrete syntax and semantics are
  obvious since they are located in the same structure. This is not the case
  in Centaur where the definition of tree building functions is required in
  addition to the abstract and concrete syntax specifications. Furthermore in
  Centaur these various specifications are given in different formalisms and
  compiled separately. Pretty-printing of abstract trees is defined in yet
  another formalism called PPML(\cite{Borras88}) while in the preceding
  example of Pecan (figure \ref{whilefig}), the source string annotated with
  SOURCE is used both for pretty-printing the syntax trees and for parsing
  typed-in text.\\

  Nevertheless, the Pecan approach is less general and the language of
  specification cannot be used to describe the features of all languages. The
  Centaur
  system is more powerful but is heavy. The
  introduction of a new language requires in practice much expertise on the
  various components of the
  system (see \cite{Rideau})\footnote{A complete language specification
  requires the use of three
  different formalisms: METAL, for the specification of concrete and abstract
  syntax (and their relationships), PPML for the pretty-printing rules and
  TYPOL for the semantic description. One can nevertheless argue that the
  introduction of a new language does not arise every day.}.\\

\subsubsection{On pretty-printing specification}

  The idea of parametrized pretty printing was developed in a very interesting
  way in the Centaur system. To build our abstract to concrete representation
  translator (unparser) we have adopted a similar approach.
  Pretty-printing is specified by annotating the production rules of the
  grammar. A general pretty printing specification is a list of
  pretty-printing rules of the form
\begin{verbatim}
  <pattern>  ->  <format>
\end{verbatim}
  where \verb+<pattern>+ is an abstract syntax tree containing variables, and
  \verb+<format>+ a formatting specification. When parsing an actual expression
  the rules are selected by matching the top of the tree to the left side of a
  production rule. When a rule is selected, variables in the pattern are
  associated with sub-trees of the tree in the order of occurrence - the
  leftmost variable with the leftmost sub-tree, the next occurrence with the
  second sub-tree from the left, and so on. On the right hand side of the rule,
  the format side, variable occurrences denote the result of a recursive call
  of the pretty printing rules on the sub-tree.\\

  For instance, in
\begin{verbatim}
(PPR) if (#cond,#stat1,#stat2) 
      -> #stat1 "if" #cond '\n'
             '\tab+' "otherwise" #stat2 
\end{verbatim}
  the occurrence of \verb+#cond+ on the left side of the arrow stands for the
  leftmost sub-tree of the tree being pretty-printed. On the right side of the
  arrow it stands for a recursive call to the pretty-printer applied to that
  sub-tree. Terms within double quotes are reserved words of the concrete
  syntax. Terms within single quotes are formatting specifications.\\

  In the Centaur system, several pretty-printing rules may match a given
  operator and a selection mechanism determines which rule obtains. The
  selection rule is simple: the rule applied is the first in the list of rule
  that matches the left-hand side. \\

  Furthermore the pretty printer takes into
  account
  the size of the displaying window and the result is very impressive. You
  can modify the
  size of a window and the text is modified so as to show all words.
  But the pattern-matching mechanism is of no help in this case and we are
  convinced that, if the result turns out to be very satisfactory, this is due
  more to the power of the formatting language than to the pattern-matching
  procedure.\\

  In our system only one rule is
  applicable at a time. Patterns can thus be viewed as abstract nodes having
  their sons as the only possible meta-variables\footnote{However this simplification
  does not result in a weaker pretty printing mechanism since the power of the
  system is determined to a larger extent by the power of the formatting
  language rather than by the way the rules are selected. Pattern
  matching only looks for inclusion of nodes. For instance, you can
  distinguish between the enclosing of an if-expression into another
  if-condition and a simple if-expression by adding a rule before the normal
  one used for the pretty printing of an {\tt if}-node:\\
  (PPR1) if (\ if
  (\#cond,stat11, stat12) ,\\
	\hspace{1.5cm} \#stat1, \#stat2 ) \\
	\hspace{0.5cm} $->$ ...\\
  (PPR2) if (\#cond,\#stat1,\#stat2) \\
	\hspace{0.5cm} $->$ \#stat1 "if" \#cond "otherwise" \#stat2 \\

  But such distinction is not obviously needed. For esthetic printing purposes,
  the
  printing properties of the sons of the current node, for example their size,
  matter most for the selection of the appropriate printing rule. These
  properties can be determined from the right side provided the language is
  powerful enough, because the recursive call to the pretty printing procedure
  appear in the right side. Therefore, the choice of the rule being selected
  is not all that important in the pretty printing mechanism. To the contrary,
  it is the formatting language which gives all its power to the system by
  means of attributes tested on the right side of the pretty-printing rule.
  Nevertheless, the pattern-matching mechanism has some advantage: it can be
  used to solve some of the problems raised by the parsing and unparsing
  of ASTs.}.\\

  As PPML (see \cite{Borras88} and \cite{Borras89}), the Absynte formalism
  describing the layout of patterns uses the notion of box. A box is either an
  atomic box or a compound box. The combination of the elements is expressed
  with box combinators and parameters.\\

  A box language has obvious advantages. In particular, it is well adapted to
  the layout of ASTs. Another interesting feature is that it can be used to mix
  text and graphic freely. The most important notion in laying out text is {\it
  indentation}. This notion covers both {\it new lines} and {\it tabulation}.
  Text always begins at the left and, as has been shown by the Centaur
  example, can be dealt with by a box meta-language that handles only
  horizontal and vertical alignments.\\

  For graphics, the situation is a little bit more complex. One would also
  like to arrange boxes in centered fashion. Secondly, one would like to print
  not only boxes but other objects as well, for example, lines between boxes.
  The way lines should be drawn should also be specified by a pretty printing
  meta-language.\\

  Both requirements are easily provided by a box meta-language. There are
  three privileged locations on an horizontal line: Left, Center and Right.
  Similarly, on a vertical line, Top, Middle and Bottom. As a result, we have
  identified six fundamental operators on boxes allowing
  to describe most of them mutual positions. \\

  To draw lines between nodes one could simply have a system which draws lines
  between the nodes in a uniform manner. For instance, it could always join
  the center of the boxes \footnote{this is the actual solution of Absynte.}.
  Another solution is to augment the meta-language and to describe the lines between a
  node and its sons. This extension does not
  necessarily requires many changes in a box language because 
  horizontal and vertical segments can be treated as boxes having a single
  dimension.\\

  Before ending this section we would like to mention a simple tool that
  we think a good programming environment should possess. One of the
  interesting feature of our text editor is its conception of templates in
  text mode. Each keyword of the language or abstract label of an abstract
  node is a possible alias for the extension of a template. Contrary to
  other editor based on the template model, the use of templates is here only
  recommended and the user can use the editor as a traditional one if he
  prefers. \\

  Two very simple generalizations of our templates mechanism could really
  accelerate the writing of programs and improve programs's comprehensibility.
  The first generalization is a good example because it shows that interesting
  tool do not necessarily require special knowledge on syntax. It consists
  in
  adding an on-line generator of aliases which adds to its aliases list any
  word typed-in by the user. In doing so, all reserved keywords will soon be
  in the aliases list, as well as all variables identifiers, type identifiers
  and user's functions names. With this facility, the user would not be afraid
  of using long function names. Suppose for instance the user has defined a
  function called CreateSimpleWindow(). He could now just type the first three
  letters, for instance Cre, followed by an escape sequence, to get the whole
  word CreateSimpleWindow be automatically inserted. If two or more aliases
  begin by Cre, the editor will offer the list of all possible choices in a
  pop-up menu. With this simple tool, programs would gain in clarity because
  long names are more precise and comprehensible. \\

  The second extension is in the same spirit. This time it is relative to
  the programming language. The idea is to recognize instructions given for
  the inclusion of modules (or librairies). Then to parse these modules or
  librairies to generate  aliases for the functions
  used in these modules. This feature would be very interesting with graphic
  librairies, because function names are in this case quite long and
  sometimes hard to remember. In particular, it would be interesting to have
  aliases for the whole definition of the functions, so as to help the user
  by giving the names and number of the arguments required.\\

\subsubsection{Parsing abstract and concrete syntax}

  Text view is a conventional view of programs. It is also one of the most
  widely used. In traditional editors, a text is treated as a list of
  characters and the compiler directly produces code from the parsing of a
  text file. In the systems where abstract syntax trees are the internal
  representation of programs, the compiler uses the ASTs representation
  directly. A text view is nevertheless proposed to cope with user's habits.
  The designer of a text view must propose a parser and an unparser to convert
  abstract syntax tree into text and {\it vice versa}.\\

  Problems are
  created by the fact that there is no {\it a priori}
  canonical transformation between abstract syntax and concrete syntax.
  Let's call for instance
  {\tt AtoC} an acceptable  transformation (from a semantic point of view) from
  abstract syntax to concrete syntax and {\tt CtoA} an acceptable converse
  transformation.  It is often difficult to
  choose a pair which satisfies the functional identity
  \begin{verbatim}
  CtoA o AtoC = Id
  \end{verbatim}
  and to preserve the original text typed-in.
  This is a major problem for the building of systems which, as Absynte,
  propose two views of the program: a pure graphic view representing an
  abstract syntax tree, and a free text view corresponding to the concrete
  syntax \footnote{Similar problems are raised by the free-text area of an
  hybrid editor.}.\\

  For instance, suppose the concrete syntax of the language allows a {\tt
  guard} statement but
  that there is only an {\tt if}-node in the AST's set. Then
  {\tt CtoA} will probably transform a guard into a cascade of if-nodes.
  But {\tt AtoC} will transform an if-node into an if-statement and then
  the cascade of if-nodes will be translated back into a cascade of concrete
  if-expressions. Thus a user who has created a piece of code as a guard in
  text mode will loose its original format when requesting the editor
  for the abstract view. \\

  The same  problem is frequently raised by more
  simple forms allowed by the concrete syntax of languages. For
  instance many languages accept that a list of type declarations be
  shortened to a single type declaration for a list of variables. This
  also creates problems with the parsing (and unparsing) of declarations as:
\begin{verbatim}
ident1, ident2: typename; 
\end{verbatim} 
which can be treated by the abstract syntax as a declarations list: 
\begin{verbatim} 
ident1: typename; 
ident2: typename;
\end{verbatim} 
  This problem arises each time the concrete syntax
  allows special abbreviations. Analogous problems
  are raised by intrinsic ambiguities of the concrete syntax. For instance if
  the conditional {\tt if}-statement allows two possible concrete
  forms - an if-then-else form and an if-then-otherwise form. Parenthesis
  around expressions also raise a similar problem. In all these cases, a
  canonical {\tt AtoC}
  transformation would by definition make a certain choice and the original
  user's representation will be lost (if not kept somewhere in the 
  internal representation of the abstract tree). \\

  Most of the systems using templates have solve part of these problems by not
  allowing all user's text representation. They also sometimes chose
  the abstract syntax so as to eliminate them. But this raises the
  following question: How should the designer of a specific language-mode
  define the abstract syntax of a language ? Can he change it at will ? If
  so, what is exactly abstract syntax ? Should abstract syntax be designed
  so as to be as close as possible to its possible concrete representations,
  or so as to minimize the required nodes for ASTs to machine-code
  transformations ?\\

  In the current version of Absynte, we do not preserve the original text
  typed-in by the user. When transformed into graphics by the graphic mode
  command, the text is first parsed and displayed as a graphic tree. If the
  tree is then modified and the user ask for displaying it as a text,
  the unparsing command (automatically generated by the concrete syntax
  specification) is launched and the editor displays its own pretty-printing
  version of the program and the previous typed-in version is lost. It could
  be possible to attach to each abstract node the rule to be applied for
  unparsing in case of ambiguities so as to maintain most of the original
  typed-in text, but this method would still failed with a few cases.\\

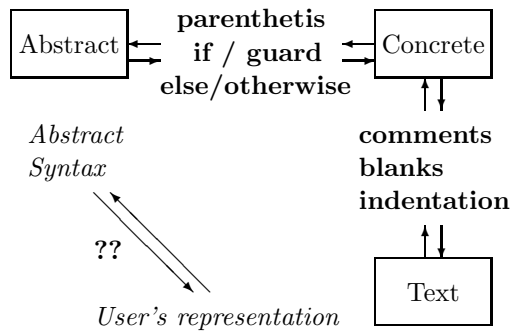
\begin{figure}[htbp]
\begin{center}
\setlength{\unitlength}{0.11mm}
\begin{picture}(620,440)(-540,40)
\put(-280,100){\vector(-1,1){120}}
\put(-100,240){\small\bf blanks}
\put(-100,200){\small\bf indentation}
\put(-100,280){\small\bf comments}
\put(-300,380){\small\bf if / guard}
\put(-320,420){\small\bf parenthetis}
\put(-340,340){\small\bf else/otherwise}
\put(-80,400){\vector(-1,0){40}}
\put(-120,380){\vector(1,0){40}}
\put(-20,360){}
\put(-80,360){\framebox(140,80){\small  Concrete}}
\put(0,360){\vector(0,-1){40}}
\put(-20,320){\vector(0,1){40}}
\put(0,180){\vector(0,-1){40}}
\put(-20,140){\vector(0,1){40}}
\put(-80,60){\framebox(140,80){\small  Text}}
\put(-340,400){\vector(-1,0){40}}
\put(-500,280){\small\it Abstract}
\put(-500,240){\small\it Syntax}
\put(-480,240){}
\put(-420,140){\small\bf ??}
\put(-420,220){\vector(1,-1){120}}
\put(-420,60){\small\it User's representation}
\put(-380,380){\vector(1,0){40}}
\put(-520,360){\framebox(140,80){\small  Abstract}}
\end{picture}
\end{center}
\caption{Parsing problems}  \label{concrete}
\end{figure}

  Other problems, besides the parsing of abstract syntax trees, are related to
  particular text token which are supposed not to be part of the concrete form
  and are already ignored by the traditional compilers. Thus elements like
  comments, tabulations and blanks, which may appear anywhere, also
  cause the failing of the \verb+"CtoA o AtoC = Id"+ equation. Figure
  \ref{concrete} classifies the problems raised by the translation of abstract
  to concrete syntax.\\

  The problem raised by comments is interesting because it underlines the fact
  that the syntax of traditional programming language has not be designed for
  human understanding (but is rather strongly machine oriented, even in the
  best cases). In a program text, comments can be added anywhere. On a tree
  representation, there are {\it a priori} only a few solutions. One can
  globally attach a comment to a node or one can attach a comment to a position
  relative to a node: before a node, onto a node and after a node. \\

  Whatever solution is retained, it will be difficult to maintain the
  coherence between the structured view and a free text view. If, for
  instance, comments are kept in the structured view by attaching a piece of
  text to a node, the cutting of this node will remove the comment while in a
  text representation the comment would have been saved. Also, if comments
  are treated as separated entities and attach to a node with an indication of
  place (as a {\tt before} annotation), the removal of the node could trigger
  a reallocation procedure determining where the comment must be replaced. But
  this new place could depend on contextual concrete features and it could
  become very difficult to automatically generate the reallocation procedure
  from traditional syntax specifications. \\

  Anyway, the fundamental question concerning comments - where to put them -
  is neither correctly answered in traditional programming. In a free program
  text for instance, some comments are actually commenting a precise
  instruction. Then they could have an equivalent on the abstract tree in
  being attached to the corresponding abstract node. But most of the time,
  comments are not attached to a precise instruction but rather to a group of
  instructions. Furthermore, if a program text contains to many comments, it
  can become (from a programmer point of view) completely unreadable. In fact,
  the program text is not a good frame for the insertion of comments.\\

  These facts have been brilliantly underlined by Knuth and partially solved
  by editors
  proposing environments for his {\it literate programming} style (\cite{}).
  The idea of a macro
  substitution of some piece of code by labeled comments templates is very
  interesting and we particularly
  appreciate it because it solves the comments problem. As a matter of fact:
\begin{enumerate}
\item It allows the hidding of uninteresting pieces in the program at 
  an arbitrary level in the tree (both for code and comments).
\item It focuses the attention on interesting pieces (both for code and
  comments) and groups them together at an arbitrary level in the tree. The
  important point here is that some parts of the tree are highlighted 
  and others are compressed without requiring that these parts be themselves
  entire sub-trees.
\item It allows a real mixing of comments in ordinary language with proper
  concrete syntax feature of the native programming language.
\end{enumerate}

  What is particularly interesting in this view of comments is that they still
  are inserted within the syntax structure but are not attached to a
  particular instruction.\\

  The reasons why such programming environments are not yet very successful
  are not to be found in their fundamental properties but in contingent
  features of their proposed implementations.  But the original idea, seen as a
  solution to the comments problem, is still valid and should be taken into
  account by future programming environment builders \footnote{WEB has the
  principal
  defect of being a Pascal's environment; it was also designed as a tool for
  the generation of a document. C-web was suffering of ergonomics problems. To
  be successful, a good implementation should integrate ideas coming from WEB
  and ideas coming from Hypertext. As in WEB, comments should be attached to
  any contiguous piece of code while preserving the original native language
  structure (as with a macro substitution). Automatic indexations should be
  used for browsing in the displayed program structure. As within an hypertext,
  the user should have access to the various components (code or
  comments or related indexed parts) with the mouse by a simple click onto the
  designed piece. Comments could then be correctly nested at any level,
  and hidden if necessary.}.\\


\subsubsection{Semantic specifications}

  A context-free grammar is generally used to describe the syntax of a
  programming language. However there are semantic features which it is not
  powerful enough to describe, e.g. the constraint that all variables must be
  declared in a program. So a more powerful specification language is
  frequently required. As well the design of specific editing functions
  frequently requires a contextual definition. For instance, we have seen that
  context sensitive functions were the only possible solution to the problem of
  program modifications in the template model.\\

  A quite popular formal approach to semantic specification uses attribute
  grammars for defining language semantics. An attribute grammar is a
  context-free grammar extended by attaching attributes to the symbols of the
  grammar. The Cornell program synthesizer is a good illustration of a
  programming environment generator built on this model. \\

  In such a system which represents programs as attributed trees, an update of
  attributes is required after each modification of the program. A general
  algorithm is to propagate changes of attribute values in the tree. The
  essential problem is that attribute grammars have non local dependencies
  among attribute values and sometimes a large number of intermediate
  attributes must be updated. So there is an efficiency problem for attribute
  updating. \\

  An optimal algorithm for incrementally evaluating attributes has been
  proposed by Reps in the Cornell program synthesizer (see \cite{Reps82},
  \cite{Reps86} and \cite{Demers81}); but this algorithm was not so fast.
  Hoover (\cite{Hoover86}) has proposed a scheme for copy bypass attribute
  propagation that dynamically replaces copy rules with non local
  dependencies, resulting in faster incremental evaluation. Other systems have
  frequently used some auxiliary data structures to record non local
  dependencies in trees (as for instance the Poe system, see
  \cite{Fischer84}).\\

  Perhaps because of the efficiency problem just mentioned, a number of
  systems have not used attribute grammars for specifying semantics. Figure
  \ref{whilefig} illustrates the method used by Pecan. The specification of
  the semantic of the {\it while} statement is provided by the annotation
  labeled SEMANTIC. This approach, using a statement-oriented semantic
  language for providing a specification is simpler than attribute grammars.
  The dependency information needed for incremental compilation is implicit in
  the language and is not separately specified. It allows almost all of the
  semantics to be provided without any external code, contrary to attribute
  systems which require semantic functions to be specially written for the
  semantic description. \\

  In Centaur, the specification of semantics uses a formalism based on natural
  deduction, called natural semantics. It is written in a language called
  TYPOL. TYPOL allows the writing of semantic rules  consisting in a
  numerator and a denominator. The numerator is collection of premisses and
  the denominator a formula which holds if all premisses of the numerator
  hold. Abstract syntax terms occur in most rules and rules can be decorated
  with lisp actions in a {\tt yacc} manner. \\

  As all METAL specifications these descriptions must be compiled and generate
  Prolog code. The compiler includes a type-checker which is  an important
  component of the system. The type-checker is written in Prolog
  because of
  the inference mechanism involved in this language. A good TYPOL
  environment exists within Centaur and many specifications have been written.
  The most important objection this system raises is that it uses much memory
  space.\\

  There are other approaches to semantic specification. For example, an
  algebraic specification formalism has also been carried out as part of
  the Centaur project. The Omega system uses a relational database to manage
  all program information. Although type-checking reduces the amount of
  handling, the problems are again efficiency and memory space.\\


\subsection{Ergonomics problems}

  The general scheme of an interface program separates internal representations
  (in our case ASTs or programs) from their external graphic representations (a
  graphic tree or a text). The external
  representation is sometimes called {\it the interface} because it provides
  an interface to internal representations. Another program called the
  interface
  is the module of interactive actions available to the user on external
  representations. Problems related to the design of this module are usually
  called {\it ergonomic} problems. Some of them are inherent to the
  structures used as external representations while others are more
  contingent and may be suppressed by taken into account a few simple rules.\\

\subsubsection{The interface design golden rules}

  To build a good interface, a designer should try as far as possible to:
\begin{enumerate}
\item increase the number of user's possible interactive actions.
\item be aware of the intrinsic {\it complexity} of interactive actions.
\item be aware that these actions be {\it naturally understood} (by the user)
\end{enumerate}

  To satisfies the first requirement both the internal
  representations and their external appearances should be interactively
  modifiable by the user. In particular, it is very useful that the user
  has the possibility of acting on external representations, not only as a
  tool for modifying internal representations but also as a means of solving
  problems raised by the external representation itself (such as its size).\\

  Another consequence of this principle is that
  a parameter used by the program and that the user
  can change in some way - such as options of command line arguments, or
  resources found in configuration startup files - should also be, as far as
  possible, {\it interactively modifiable} (i.e. on-line) by the user.
  Frequently, this generalization will not increase the cost of software
  development.\\

  In Absynte for instance, the mode of graphic representation (vertical
  centered, horizontal centered or simple shown in figure \ref{screen}) can be
  changed within the mode menu. 
  Besides all these different modes
  have proper parameters - vertical and horizontal spacing between boxes -
  that can also be changed interactively. Text windows,
  tabulation sizes, fonts displaying text and aliases definitions for
  templates can also be changed interactively by the user. These features give
  more flexibility and are usually well appreciated by users.\\

  The two following recommendations should also be taken seriously by interface
  builders to increase the power of their interface. Just as a degree of
  complexity can be associated to an algorithm, one can associate to an
  interactive action a certain pragmatic cost by counting the number of
  physical actions required to perform this action (drag the mouse pointer,
  click on a button, change the focus from the mouse to the
  keyboard or reverse, etc.). Likewise the cost of its {\it natural
  understanding}
  can be calculated by counting the number of conventions the user must
  learn in order to be able to perform this action.\\

  A good interface for beginners should particularly minimize the later cost.
  But to satisfies all users, the first one should also be reduced.
  Unfortunately these two requirements generally conflict. For instance, the
  launching of a command from a menu in a title bar does not require a special
  knowledge because it is based on a natural convention. But on the other
  hand, it is a complex action. It actually
  requires: a possible change of focus (keyboard to mouse), a possibly
  long dragging of the mouse pointer to the menu bar, a click on the title
  menu item, a short dragging of the mouse to the required item command,
  a new click (or release) on it, and another possible change of focus (mouse
  to keyboard).\\

  Another example illustrating this point is given by pop-up menus. Pop-up
  menus (appearing anywhere in a window) are less costly than menus in a
  title bar, but they require the learning of an action to trigger the pop-up.
  Thus their complexity decreases while their cost of comprehensibility
  increases\footnote{Such considerations were part of the motivation for the
  definition of interface standards like OSF/MOTIF or OPENLOOK. But standards
  emerge usually {\it de facto} and it is rather difficult to impose them.
  What can be said to-day is that a few standard objects have emerged:
  menu-bars, scrollbars, cancel and confirm buttons in dialogue boxes, text
  cursors, cut/copy/paste conventions, an the like.\\

  The Apple MacIntosh interface is in a certain sense the actual standard.
  Unfortunately, this model cannot be directly adapted to other graphic
  workstations. For instance, the cut/copy/paste command work between all
  applications on a MacIntosh because these applications have been built on
  the same machine. Besides, you probably know where the help for any
  application is on a MacIntosh (it is in the apple).}.\\


  To solve such problems a graphic interface can provide two ways for launching
  the most frequent commands. One way will minimize the cost for understanding
  and the second way will minimize its practical complexity. A good solution
  is to provide for each command accessible through a menu bar a keys's
  combinaison to launch it from anywhere else.\\

  To the preceding recommendations on the design of interactive actions should
  also be added the following rules of dialog design taken from Ben
  Shneiderman in \cite{Schne}:
\begin{enumerate}
\item {\it Strive for consistency.}
  The consistency principle is the most frequently violated one, and yet the
  easiest one to repair and avoid. Consistent sequences of actions should be
  required in similar situations, identical terminology should be used in
  prompts, menus and help windows, and consistent commands should be employed
  throughout. 
\item {\it Offer simple error handling.}
   As much as possible, design the system so the user
  cannot make a serious error. If an error is make try to make the system
  detect the error and offer simple, comprehensible mechanisms for handling
  the error.
\item {\it Offer informative feedback.}
  For every operator action there should be some system feedback. For frequent
  and minor actions the response can be very modest, whereas for infrequent
  and major actions the response should be substancial.
\item {\it Permit easy reversal of actions.}
  As much as possible, actions should be reversible. This relieves anxiety
  since the user knows that errors can be undone, and encourages exploration
  of unfamiliar options.
\end{enumerate}
  
\subsubsection{On graphic representations}

  Other ergonomics problems we have encountered were due to our external
  representations themselves. They were related to the view of programs as
  graphic trees.
%
  With graphic trees, nodes naturally extend vertically. If
  the height of your terminal is too small, you can try to decrease the
  vertical spaces between boxes. But in doing so, you will quickly use all the
  available space. Worse, usually the width is exhausted before the height. So
  you simply cannot do anything but scrolling. Another solution is to change
  the size of the primitive elements (as the zoom operation of the MacIntosh).
  But it is very expensive, because it requires changing the fonts
  and compute the entire display all over again. \\

  In addition to scrolling and zooming, we have proposed another feature to
  solve this problem. We have introduced
  the notion of {\it compressed} tree-node. A {\it compressed} node is a node
  which has the same behavior than an icon. It is represented as a simple gray
  node - even if it has many sons. When a clicking on it occurs, it opens as a
  window, showing the whole tree, father and sons, in text or graphic mode.\\

  So, for each node in the tree, it is possible to display it using the whole
  of the screen. Furthermore, the notion of {\it compressed} node
  introduces also a way of mixing text and graphic since in the window
  attached to a node, the tree can be displayed in text or graphic mode.\\

  Nevertheless this problem - usually referenced as the elision problem - is
  never fully satisfactory solved. Here the intrinsic difficulty is that,
  independently of  practical considerations such as the graphic size of a
  tree, the user would like to use
  elision as a semantic tool, to hide  uninteresting parts of the program and
  not only as a tool to solve a space problem.\\

  In text mode the elision problem is easier to solve. As we have mentionned
  in section 2.2.3. a special comments's handler, based on Knuth's idea, but in
  the spirit of hypertext, could be a very attractive feature. Another simpler
  solution is to provide a way of hiding the body of functions - while only
  showing their declarations. Note that this solution, to hide the body of
  functions, is not a solution for a graphic tree representation. First
  because it would lead to a very desequilibrated tree if only one function is
  expanded and secondly because most of the time, the function itself would
  be to big to be usefully displayed on the screen.\\

  Our present conclusion is that graphic trees are pretty but are really not
  {\it ergonomic} and really inappropriate for program editing. The main
  reason is that, probably because of their cognitive habilities, humans
  cannot easily read graphs, and, in particular,
  they get not really ready for understanding trees \footnote{This is a general
  problem with graphic representations. For experienced users,a tabular
  display of fifty document names may be more appealing than only ten document
  graphic icons with their names abbreviated to fit the icon size.}.\\

\begin{figure}[htbp]
\begin{center}
\setlength{\unitlength}{0.11mm}
\begin{picture}(300,500)(-300,20)
\put(-220,80){\framebox(140,40){}}
\put(300,200){\framebox(0,20){}}
\put(-220,140){\framebox(140,40){}}
\put(-80,200){\framebox(0,20){}}
\put(-220,200){\framebox(140,40){}}

\put(-240,60){\framebox(180,240) [tl] {\small    case}}
\put(120,160){\framebox(0,20){}}
\put(-240,320){\framebox(180,120)[tl] {\small    if}}
\put(-280,40){\framebox(260,460) [tl] {\small    program}}
\end{picture}
\end{center}
\caption{A new text boxes style}  \label{word}
\end{figure}
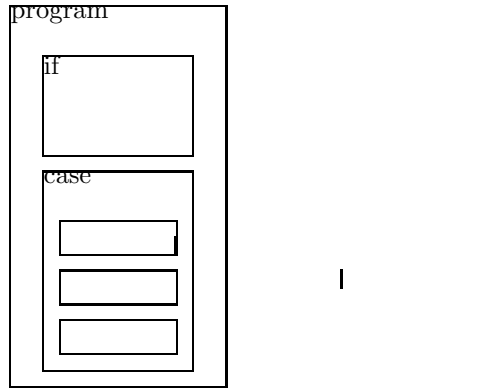

  The best solutions for syntax directed-editors based on
  ASTs is to display them in text form, with a few features showing the
  internal tree structure. Our current preference is for the style 
  of Word 4 on the Apple MacIntosh which allows the selection and move of
  boxes (enclosing an indentation text level) in the so-called {\it outlining}
  mode of display. We have not yet begin the implementation of this new style,
  but figure \ref{word} gives the flavour of it.

\section{Conclusion}

  In this paper we gave a detailed analysis of 
  the use of abstract syntax trees in the development of programming
  environments. Our main conclusion is that, by their structural
  aspect, ASTs apparently facilitate language specifications and lead to the
  development of multi-language systems allowing the generation of
  syntax-directed editors. But these programming environments (most of the
  time based on the template model) are not efficient enough to satisfy the
  user. Thus designers are lead to build hybrid editors, allowing
  free editing areas for typing program text as in traditional text editors.
  Then the original property of ASTs - the removal of concrete syntax - is
  partially lost and new problems are raised.\\

  What were the goals of structured editing ? Mainly accelerate the process of
  software production and help the programmer in understanding its program.
  But what we have observed in using our abstract syntax tree editor is that
  the granularity of the syntax is never well suited to provide a good
  representation of a piece of code (the only exception being may be the case
  statement). Most of the time, it is too fine - as for instance near the
  leaves of the tree (think to an arithmetic expression), and conversely 
  at some more higher level - such as functions definition - the
  structure is entirely flat (in most programming languages, it is a flat list
  of declarations). \\

  Syntax directed editors are only tools helping the programmer to understand
  the syntactic structure of a program. For this reason, they are of no help
  for
  an experimented programmer who is very familiar with the syntax. Suppose you
  have a syntax editor for English, showing and correcting syntax errors. Such
  an editor is certainly very useful for foreigners but not so much for
  English native speakers. A good word processor, providing spelling but also
  automatic table of contents generation and index generation, outlining
  features, etc., is really more appreciated by English writers \footnote{A
  good question is what could be the equivalent for programming ?}.\\

  Traditionally, syntax has been opposed to semantics. The meaning of a program
  is supposed to be given by its text and emerge from its syntactic structure.
  In denotational semantics, a program denotes an abstract function which can
  be computed from the meaning of its syntactic components - the basic
  component being the statement. In natural languages, the meaning of a
  sentence cannot be entirely computed from its syntactic structure. There are
  other levels - called {\it pragmatic} \footnote{Concepts used in pragmatics
  are for instance speaker's intention, common knowledge, inference,
  relevance. They are used to explain how people do actually communicate.} -
  which must be taken into account to understand what the speaker said when
  uttering a sentence. \\

  We think that the same kind of distinction could be made as far as the
  communication of programs is concerned. The present programming framework
  has been build for the communication of a program from a programmer to a
  machine. It does not really allows the communication of a program from a
  programmer to another programmer. In this framework no real common database
  of programs can be build and programming stays an obscure individual task.\\

  Large software programs are inherently complex, and there is no royal road
  to instant comprehension of their features. But one thing is obvious: there
  is a real need of another structured level to fragment a program in more
  pieces. Because no language or programming environment provides it to-day
\footnote{This is not absolutely true. Object oriented languages are a first 
	step in this direction. But although they provide another level
  	of structuration by means of objects, the general architecture
  	is de facto hidden because of the general control mode. Specifications
  	environments also provide a frame for this task, but in our opinion,
  	the specification task should not be performed separately.
}, programmers tend to use the file system for this purpose. It is symptomatic
  that big programs are divided into separate files within several
  directories. Here the names chosen for the directories and files are of
  major importance. "Readme" files and "makefile" also help in
  understanding the organization of the program. Systems
  functions, like {\tt ctags} or {\tt grep} are also very used to
  recover pragmatic links between variables defined within the various
  files.\\

  To clarify the pragmatic meaning of a program, one should introduce some
  higher-level structures and give more importance to modules and their
  possible relationships. A good programming environment should have
  facilities allowing the user to have a structural view of the program. To
  pursue the word processor's metaphor, some kind of table of contents should
  be displayed on the screen, to give the user a global view of the
  program and to provide a way to point at a module or a function in
  order to edit its content. It is not very clear whether this table of
  contents should be organized as a linear structure (as it is with text
  processing) or as a tree or a graph. This is an open question. In a
  persistent framework, this question should also
  be related to the questions concerning the organization of the persistent
  database itself. \\

\end{document}